\documentclass[preprint,prl,superscriptaddress,prbib,11pt]{revtex4-1}
\usepackage{graphicx}
\usepackage{bm}
\usepackage{hyperref}
\usepackage{todonotes}
\usepackage{nicefrac}
\usepackage{bbm}
\usepackage[normalem]{ulem}
\usepackage{booktabs}
\usepackage{subfigure}
\usepackage{rotating}
\usepackage{color}
\usepackage{float}
\usepackage[T1]{fontenc}
\usepackage[utf8]{inputenc}

\begin{document}

\title{Broadband nonlinear Hall response and multiple wave mixing in a room temperature altermagnet}

\author{Soumya Sankar}
\thanks{These authors contributed equally.}
\affiliation{Department of Physics, The Hong Kong University of Science and Technology, Clear Water Bay, Kowloon, Hong Kong SAR}

\author{Xingkai Cheng}
\thanks{These authors contributed equally.}
\affiliation{Department of Physics, The Hong Kong University of Science and Technology, Clear Water Bay, Kowloon, Hong Kong SAR}

\author{Xinyu Chen}
\thanks{These authors contributed equally.}
\affiliation{Department of Physics, The Hong Kong University of Science and Technology, Clear Water Bay, Kowloon, Hong Kong SAR}

\author{Xizhi Fu}
\affiliation{Department of Physics, The Hong Kong University of Science and Technology, Clear Water Bay, Kowloon, Hong Kong SAR}

\author{Takahiro Urata}
\affiliation{Department of Electrical, Electronic and Computer Engineering, Gifu University, Gifu, Gifu 501-1193, Japan}

\author{Wataru Hattori}
\affiliation{Department of Materials Physics, Nagoya University, Nagoya, Aichi 464-8603, Japan}

\author{Wenlong Lu}
\affiliation{Department of Physics, The City University of Hong Kong, Kowloon, Hong Kong SAR}

\author{Zihan Lin}
\affiliation{Department of Physics, The City University of Hong Kong, Kowloon, Hong Kong SAR}

\author{Dong Chen}
\affiliation{College of Physics, Qingdao University, Qingdao 266071, China}

\author{Claudia Felser}
\affiliation{Max Planck Institute for Chemical Physics of Solids, 01187 Dresden, Germany}

\author{Hiroshi Ikuta}
\affiliation{Department of Materials Physics, Nagoya University, Nagoya, Aichi 464-8603, Japan}
\affiliation{Research Center for Crystalline Materials Engineering, Nagoya University, Chikusa-ku, Nagoya, Aichi 464-8603, Japan}

\author{Junzhang Ma}
\affiliation{Department of Physics, The City University of Hong Kong, Kowloon, Hong Kong SAR}

\author{Junwei Liu}
\email{liuj@ust.hk}
\affiliation{Department of Physics, The Hong Kong University of Science and Technology, Clear Water Bay, Kowloon, Hong Kong SAR}

\author{Berthold J\"ack}
\email{bjaeck@ust.hk}
\affiliation{Department of Physics, The Hong Kong University of Science and Technology, Clear Water Bay, Kowloon, Hong Kong SAR}

\date{September 2025}

\begin{abstract}
%~200 words
Crystalline symmetries determine the linear and nonlinear response of materials to external stimuli such as mechanical pressure and electromagnetic fields, governing phenomena such as piezoelectricity, optical activity, and multiple wave mixing with wide ranging technological applications. Altermagnets present a new class of materials with magnetic crystalline order where specific crystal symmetry operations connect antiferromagnetic sublattices, leading to non-relativistic spin-splitting of the electronic band structure. Hence, the electric material properties of altermagnets should uniquely mirror these underlying symmetry properties, potentially giving rise to novel phenomena in response to external driving fields. Here, we report the discovery of a broadband third-order nonlinear anomalous Hall effect in altermagnetic CrSb at room temperature. The comparison of our observations with symmetry analyses and model calculations shows that this nonlinear Hall response is induced by the nonlinear electric susceptibility of a Berry curvature quadrupole, which exists within the spin-split band structure of CrSb and is characterized by the underlying crystalline and magnetic symmetries. We then utilize this third-order nonlinear electric susceptibility of CrSb to realize a multiple wave mixing device with pronounced four wave mixing output, which could, in principle, be extended to THz frequencies. Our study discovers that the crystalline magnetic order of altermagnets determines their nonlinear electric material properties, which could facilitate applications in high-frequency electronics, THz generation, communication networks, and energy harvesting.
\end{abstract}

\maketitle

\section{Introduction}
\begin{figure}[H]
    \centering
    \includegraphics[width=1\textwidth]{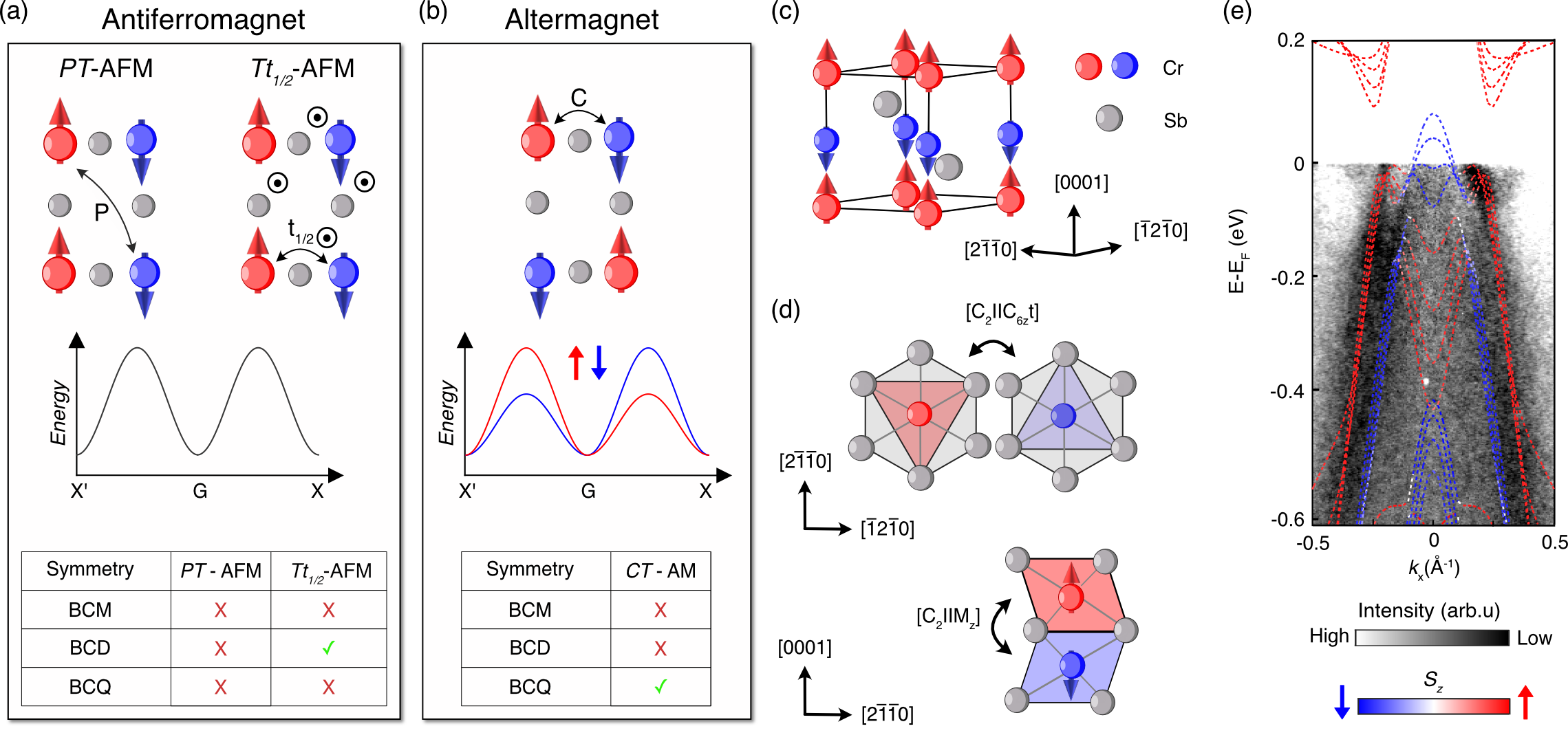}
    
     \caption{{\bf CrSb: a g-wave altermagnet.} (a) Top left: Shown is the lattice structure of a $\mathcal{P}\mathcal{T}$-symmetric antiferromagnet, where the atomic sublattices are connected via inversion ($\mathcal{P}$) and time-reversal ($\mathcal{T}$) symmetry. Up and down spin orientations are color-coded in red and blue color, respectively. Top right: Shown is the lattice structure of a t$\mathcal{T}$ antiferromagnet, where the inversion is broken due to the presence of an extra grey atom out of the plane. Center: Shown is the schematics of the spin-degenerate electronic band structure of both $\mathcal{P}\mathcal{T}$ and t$\mathcal{T}$ antiferromagnets within the Brillouin zone. Bottom: The table illustrates which Berry curvature contributions are allowed (green check) and forbidden (red cross) to exist in antiferromagnets by symmetry arguments. BCM, BCD, and BCQ denote the Berry curvature monopole, dipole, and quadrupole, respectively. (b) Top: Shown is the illustration of the lattice structure of an altermagnet, schematically indicating the presence of $C$-symmetry operation to connect the two spin sublattices. Center: Shown is a schematic illutstration of the spin-split electronic band structure of an altermagnet within the Brillouin zone. Bottom: The table illustrates which Berry curvature multipoles are allowed to exist in altermagnets by symmetry arguments. (c) Shown is the crystallographic and magnetic structure of CrSb. The Chromium (Cr) atoms with up-spin and down-spin are shown in red and blue color, respectively and antimony (Sb) atoms are shown as grey spheres. (d) Top: Shown are the anisotropic spin sublattices of CrSb connected via the combined symmetry rotation operation $[C_{2}^s {\parallel} C_{6z}t]$. Bottom: Shown are the anisotropic spin sublattices of CrSb connected via the mirror operation $[C_{2}^s {\parallel} M_{z}]$. (e) Shown is the momentum $k_x$ and energy $E$-resoled angle-resolved photo-electron spectroscopy spectrum recorded on the surface of a cleaved CrSb crystal at $k_y= 3.3\,Å^{-1}$ and $k_z=1.2\,Å^{-1}$ along a direction parallel to $\Gamma-K$ (see Methods section for details).}
    \label{fig:fig1}
\end{figure}

The response of a material to external stimuli, such as electric and magnetic fields, light, and mechanical pressure, is often described by the susceptibility $\chi$ whose properties are governed by the underlying crystalline symmetries. The piezoelectric effect in non-centro-symmetric crystals and circular dichroism of materials with chiral symmetry, as well as multiple wave mixing are examples of linear and nonlinear material responses that lead to wide-ranging technological applications. Altermagnets are a new class of antiferromagnets whose magnetic structure is uniquely connected to crystalline symmetries~\cite{hayami2020bottom, ma2021multifunctional, vsmejkal2022beyond}. In conventional antiferromagnets, the spin-up and spin-down lattices are connected by half-translation ($t_{1/2}$) or inversion ($\mathcal{P}$) symmetry, which guarantees the spin degeneracy of the electronic states in momentum space, as illustrated in Fig.~\ref{fig:fig1}(a). In contrast, in altermagnets, crystalline anisotropy results in spatially anisotropic spin-up and spin-down sublattices. These sublattices are connected by rotation or mirror transformations that break $\mathcal{P}\cdot \mathcal{T}$ and $\mathcal{T}\cdot t_{1/2}$ symmetries, where $\mathcal{T}$ denotes the time-reversal symmetry, resultting in non-relativistic spin-splitting of the electronic band structure~\cite{krempasky2024altermagnetic, ding2024large, reimers2024direct, zhang2025crystal,jiang2025metallic}, as schematically illustrated in Fig.~\ref{fig:fig1}(b). Hence, altermagnetic order can be regarded as a magnetic crystal order that is characterized by the crystalline symmetries and the direction of the magnetic order (Néel) vector~\cite{wu2007fermi,vsmejkal2022anomalous,hu2025catalog,Zhou2025Manipulation}. Consequently, the presence and specific type of altermagnetic order parameter should in general determine the susceptibility of altermagnets to an external driving field.

%{[\bf P2]} The response of a metal to an externally applied electric field is characterized by the conductivity tensor $\sigma$ whose diagonal tensor elements describe the longitudinal, or in other words conventional, electric conductivity. In conventional metals, the application of an external magnetic field or the presence of ferromagnetism along a direction $\hat{\bf n}$ causes a Hall response, which manifests as a transverse current and is described by the off-diagonal elements of $\sigma$. This conventional Hall effect has no symmetry requirements other than broken $\mathcal{T}$ and thus is isotropic in a plane with surface normal direction $\hat{\bf n}$. On the other hand, crystalline symmetries can permit higher-order contributions to the conductivity matrix that arise from the quantum-geometrical tensor of the electronic states, such as Berry curvature and quantum metric~\cite{gao2014field, sodemann2015quantum, zhang2023higher}. These higher order tensor elements give rise to nonlinear anomalous Hall effects (NLAHE) that are anisotropic in real space and determined by the underlying crystal symmetries~\cite{ma2019observation, lai2021third, gao2023quantum, wang2023quantum, sankar2024experimental}. 

It has been proposed that the magnetic crystalline order of altermagnets can support Berry curvature multipoles, arising in the non-relativistic spin-split electronic band structure~\cite{fang2024quantum, ezawa2024intrinsic}, which can give rise to nonlinear anomalous Hall effects (NLAHE)~\cite{ma2019observation, lai2021third, gao2023quantum, wang2023quantum, sankar2024experimental}. When the crystalline structure of altermagnets preserves inversion symmetry and the magnetization is compensated ${\bf M }=0$, Berry curvature monopoles (BCM) and dipoles (BCD) are prohibited by symmetry, but a Berry curvature quadrupole (BCQ) can generally appear, as shown in Fig.~\ref{fig:fig1}(b). Therefore, altermagnets could exhibit a nonlinear susceptibility $\chi^{3\omega}$ in their material properties induced by a BCQ, potentially giving rise to nonlinear effects, such as the electric third-order NLAHE. This property qualitatively distinguishes altermagnets from $\mathcal{P}\mathcal{T}$-symmetric and $\mathcal{P}$-breaking antiferromagnets, which cannot exhibit a BCQ on symmetry grounds, as summarized in Fig.~\ref{fig:fig1}(a). 

Moreover, nonlinear responses of altermagnets as a manifestation of an intrinsic nonlinear susceptibility, such as third-order NLAHE, will exhibit a strong spatial anisotropy, determined by the symmetries of the underlying magnetic crystal order.
Hence, they can be qualitatively distinguished from anomalous Hall effects in the linear response that, while prohibited by symmetry in altermagnets, could be induced by relativistic spin-orbit coupling~\cite{feng2022anomalous, gonzalez2023spontaneous, kluczyk2024coexistence, chilcote2024stoichiometry, mazin2024origin}. To date, concepts to utilize the spin-split band structure of altermagnets mostly focus on spintronic applications that remain limited to materials with $d-$wave order parameter by symmetry constraints~\cite{bose2022tilted, bai2022observation, karube2022observation, gonzalez2021efficient, bai2023efficient, shao2023neel, zhou2024crystal, badura2025observation}. Therefore, experimental insight into the general impact of the non-relativistic spin splitting on the linear and non-linear electric material properties of altermagnets will be critical to advance our fundamental knowledge and identify technological application scenarios of this new class of magnetic materials.

In this work, we provide experimental evidence for a broadband nonlinear anomalous Hall response in a room temperature altermagnet. Conducting electric Hall effect measurements of altermagnetic CrSb along different spatial directions, we detect a distinct third-order NLAHE. The comparison with symmetry analyses and model calculations makes a compelling case that this effect is induced by a Berry curvature quadrupole, which exists within the non-relativistic spin-split electronic band structure in a perfect altermagnet without any additional net magnetic moment. We also show that the broadband nonlinear susceptibility $\chi^{3\omega}$ of CrSb can be utilized to realize a multiple wave mixing device at room temperature, expanding the potential application scope of altermagnets beyond spintronics to high-frequency electronics, communication networks, and energy harvesting~\cite{du2021nonlinear}. 

\section{Symmetry properties of altermagnetic order in C\lowercase{r}S\lowercase{b}}

CrSb crystallizes in a hexagonal lattice structure ($a = 4.12\,$Å, $c = 5.47\,$Å, $P6_3/mmc$ space group)~\cite{takei1963magnetic}. Its antiferromagnetic structure can be characterized by an A-type Néel vector whose easy axis points along the crystallographic $c-$axis~\cite{snow1952neutron, abe1984magnetic}, as seen in Fig.~\ref{fig:fig1}(c). In this configuration, the magnetic moments of Cr pointing along the $c-$axis are aligned in parallel within the hexagonal Cr layer and antiparallel between two neighboring layers. For simplicity, we define a Cartesian coordinate system in which the $x$, $y$, $z$ directions correspond to the crystallographic $[2\bar{1}\bar{1}0]$, $[01\bar{1}0]$, and $[0001]$ directions, respectively. In the absence of relativistic spin-orbit coupling (SOC), which is generally weak for Cr $3d$ electrons, the magnetic structure of CrSb can be described by spin-group formalism~\cite{brinkman1966theory, litvin1974spin, vsmejkal2022beyond, chen2024enumeration, jiang2024enumeration, xiao2024spin}. Within this notation, the spin sublattices of CrSb are connected by two symmetry operations $C_2^s M_z$ and $C_2^s\tilde{C}_{6z}$, where $C_2^s$, $M_z$, and $\tilde{C}_{6z}=C_{6z}t_{1/2}$ denote a spin-flip operation, mirror-z operation, and six-fold out-of-plane rotation $C_{6z}$ combined with a half-translation $t_{1/2}$ along [0001], as schematically illustrated in Fig.~\ref{fig:fig1}(d). Hence, CrSb can be classified as a bulk $g$-wave altermagnet~\cite{vsmejkal2022beyond} whose magnetic structure breaks $\mathcal{P}\cdot\mathcal{T}$ and $\mathcal{T}\cdot t_{1/2}$. The breaking of these combined symmetries manifests in non-relativistic spin-splitting of the electronic states in momentum space~\cite{reimers2024direct, ding2024large, yang2025three, lu2025signature}, a characteristic feature of altermagnetism.

We visualize this momentum-space splitting of the electronic bandstructure of CrSb by conducting angle-resolved photo-electron spectroscopy (ARPES) measurements on the surface of cleaved bulk crystals (see Methods section for details). The symmetry transformation $C_2^s M_z$ and $C_2^s\tilde{C}_{6z}$ between the spin sublattices manifest in nodal planes of spin-degenerate bands along high-symmetry directions and planes ~\cite{vsmejkal2022beyond, yang2025three}. Therefore, to visualize the spin-split band structure, we have performed a spectroscopic cut away from high-symmetry lines at $k_y=3.3\,Å^{-1}$ and $k_z=1.2\,Å^{-1}$ along a direction parallel to $\Gamma-K$ (see Sec.~I for details of the cut location within the $xy-$plane of the Brillouin zone). The resulting energy-dependent ARPES signal is shown in Fig.~\ref{fig:fig1}(e). As can be seen, our experimentally detected signal is in good agreement with the spin-split electronic band structure obtained from DFT calculations, which is overlaid to the data (see Methods section for details). We detect the non-relativistic band splitting of the bands down to $0.2\,$eV below Fermi energy, consistent with recently published results~\cite{reimers2024direct, ding2024large, yang2025three, lu2025signature}.

In the next step, we derive a qualitative understanding of the linear ($n=1$) and nonlinear ($n>1$) current responses ${\bf j}^{n\omega}={\sigma}^{n\omega}({\bf E}^{\omega})^n$ of CrSb to the application of an a.c. external electric field ${\bf E}^{\omega}$ of frequency $\omega$. These responses are determined by the linear and nonlinear electric conductivity tensors $\sigma^{n\omega}$. In simple terms, $\sigma^{n\omega}$ accounts for electric material properties, such as the electric resistivity and the electric Hall effect. Interestingly, the quantum-geometric tensor $\mathcal{Q}$ of the electronic wavefunction, which is determined by all crystalline and magnetic material symmetries, can also contribute to $\sigma^{n\omega}$~\cite{gao2014field, sodemann2015quantum, zhang2023higher}. The Berry curvature $\bf \Omega$, which corresponds to the imaginary part of $\mathcal{Q}$, can contribute to off-diagonal elements and manifest in the observation of linear and nonlinear anomalous Hall effects. The quantum metric, which corresponds to the real part of $\mathcal{Q}$, can generally contribute to all tensor elements of the nonlinear conductivity tensors ($n>1$) and give rise to both longitudinal and transverse responses. Note that the quantum metric does not contribute to the linear conductivity tensor ${\bf\sigma}^{1\omega}$.

We can derive specific contribution of $\mathcal{Q}$ to the tensor elements of $\sigma^{n\omega}$ for $n\geq1$ in CrSb by conducting a symmetry analysis. Details of the symmetry analysis are shown in Sect.~II of the Suppl.~Materials and summarized in Fig.~\ref{fig:fig2}(a). The altermagnetic structure of CrSb breaks the combined $\mathcal{P}\cdot \mathcal{T}$ symmetry and generally allows for a finite $\bf\Omega$ in the Brioullin zone. However, when spin-orbit coupling is weak, CrSb exhibits compensated magnetization ${\bf M}=0$ and thus preserves $C_2^s M_z$ and $C_2^s\tilde{C}_{6z}$ symmetry. This will force integration of $\bf\Omega$ over the Brillouin zone to vanish and  lead to the absence of the first-order anomalous Hall conductivity, $\sigma_{\beta\alpha}^{1\omega}$, where $\alpha$ and $\beta$ denote spatial directions that are perpendicular to each other and lie within one plane. Therefore, by symmetry arguments, $\sigma_{\beta\alpha}^{1\omega}=0$ and altermagnets do not exhibit an intrinsic first-order anomalous Hall effect (AHE) induced by the non-relativistic spin-splitting of the electronic structure. Regarding the second order nonlinear conductivity tensor ${\bf\sigma}_{2\omega}$, Berry curvature and quantum-metric contributions require broken $\mathcal{P}$. However, the centro-symmetric lattice structure of CrSb preserves $\mathcal{P}$, and hence ${\bf\sigma}^{2\omega}=0$ and CrSb are expected to exhibit no longitudinal or transverse second-order response.

Unlike the first- and second-order responses, our analyses show that the magnetic crystal order of CrSb does support a BCQ and a nonlinear third-order response. Specifically, we find that only one quadrupole $\mathcal{Q}_{xxxz}$ exists, which lies within the $xz-$plane of the Brillouin zone, and contributes to exactly one off-diagonal component $\sigma_{zxxx}^{3\omega}$ of the third-order conductivity tensor, as seen in Fig.~\ref{fig:fig2}(a). Hence, CrSb should exhibit a third-order nonlinear susceptibility $\chi_{zxxx}$, which can give rise to a third-order NLAHE induced by a BCQ when an a.c. current of frequency $\omega$ is applied along the $x$-direction of the crystal. This NLAHE would manifest as a finite transverse voltage response $V_{zxxx}^{3\omega}$ at the third harmonic $3\omega$ along the $z-$direction. Critically, owing to the anisotropic nature of the BCQ distribution in the Brillouin zone, CrSb should not exhibit a third-order Hall response when the current is applied along the $y-$ or $z$-directions, that is, $V_{zyyy}^{3\omega}=0$ and $V_{xzzz}^{3\omega}=0$, respectively. Meanwhile, other potential contributions to $\sigma_{zxxx}^{3\omega}$, such as quantum metric and Drude effects, are forbidden by the crystal symmetry of CrSb. 

The absence of first- and second-order responses and the predicted third-order nonlinear suscpetibility $\chi^{3\omega}$ of CrSb represent unique intrinsic properties of altermagnetic order in the non-relativistic limit and the spatial anisotropy of $\chi^{3\omega}$ precisely reflects the magnetic and crystalline symmetries encoded in the altermagnetic g-wave order parameter of CrSb~\cite{vsmejkal2022beyond}. Finally, also note that our symmetry analysis also shows that CrSb can host quantum metric quadrupoles (QMQ), which contribute to the diagonal-elements $\sigma_{\alpha\alpha\alpha\alpha}^{3\omega}$, that is, the longitudinal voltage responses along all crystallographic directions, as summarized in Fig.~\ref{fig:fig2}(a). Critically, the presence of finite QMQ tensor elements does not have specific symmetry requirements. Hence, QMQ-induced longitudinal voltage responses are not a distinguishing feature of altermagnets as they can generally appear in all magnetic configurations (ferromagnetic, antiferromagnetic, altermagnetic, and non-magnetic).

\section{Observation of the third-order NLAHE at room temperature}

\begin{figure}[H]
    \centering
    \includegraphics[width=1\linewidth]{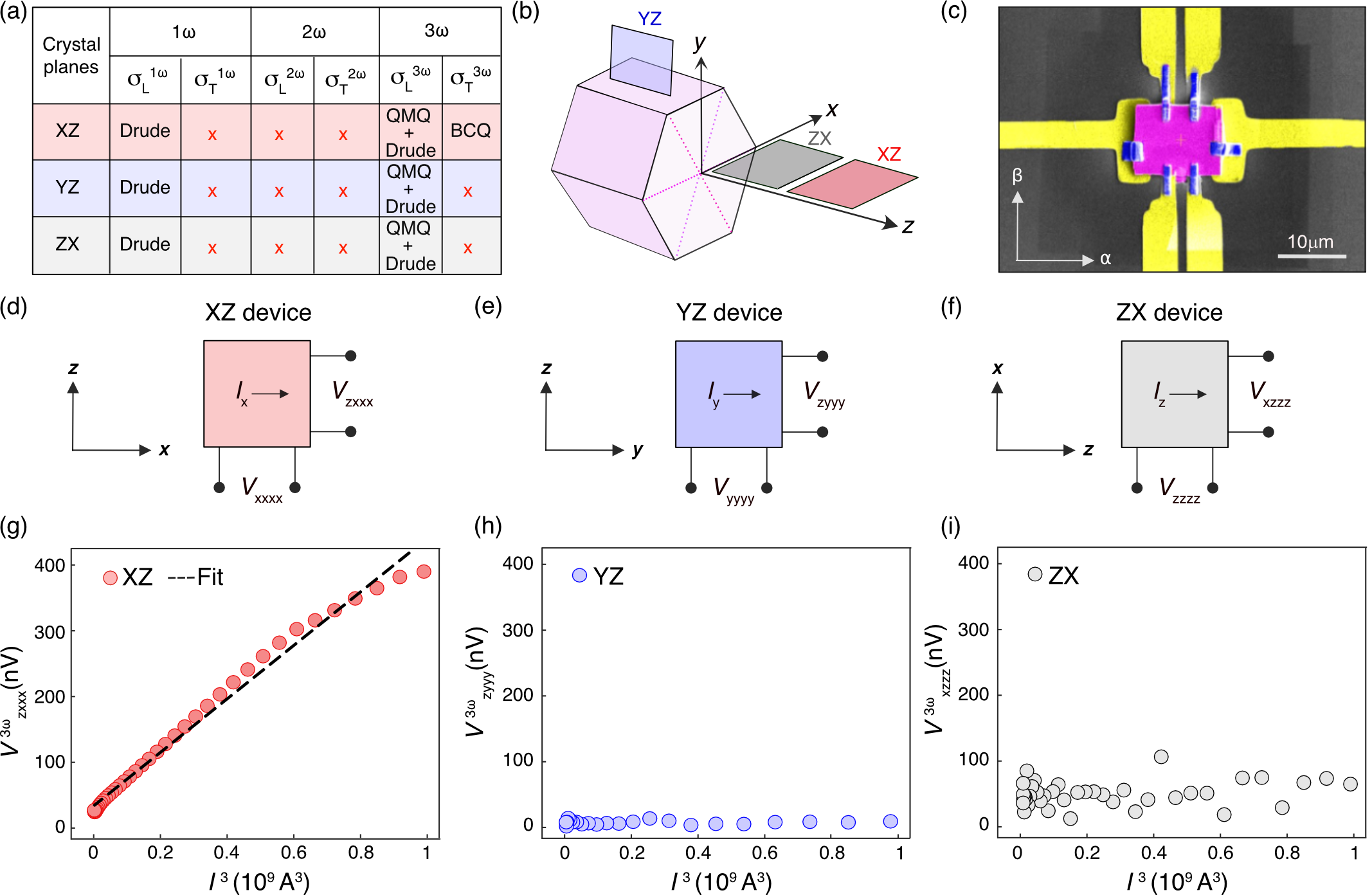}
    \caption{\textbf{Discovery of the third-order nonlinear anomalous Hall effect in altermagnetic CrSb at room temperature.} 
    (a) Shown is a summary of theoretically possible contributions--Drude scattering, quantum metric quadrupole (QMQ), and Berry curvature quadrupole (QMQ)--to the longitudinal ($\sigma_{\rm L}^{n\omega}$) and transverse ($\sigma_{\rm T}^{n\omega}$) linear ($1\omega$) and nonlinear ($2\omega,\,3\omega$) conductivities, respectively obtained from symmetry analysis within different crystallographic planes of CrSb. Red crosses indicate that contributions are prohibited by symmetry arguments.
    (b) Shown is a schematic illustration of the prism-shaped CrSb bulk crystal (light purple) as well as the spatial orientation of the lamellae ($'XZ'$ red, $'YZ'$ blue, and $'ZX'$ grey) with respect to the bulk crystal.
    (c) Shown is a false-colored scanning electron microscopy image of the finished Hall bar device $'XZ'$ made from a crystalline lamella cut within the crystallographic $xz-$plane. The lamella, FIB-deposited Pt contacts, and prepatterned gold electrodes are highlighted by pink, blue, and yellow color, respectively. The SiO$_{2}$/Si substrate appears as black background.
    (d)-(f) Shown are schematic illustrations of the measurement geometry of the $'XZ'$, $'YZ'$, and $'ZX'$ devices, respectively. The bias current $I_{\alpha}$ and the longitudinal $V_{\alpha\alpha\alpha\alpha}$ and transverse $V_{\beta\alpha\alpha\alpha}$ voltage responses are denoted within the respective coordinate frame of each device.
    (g)-(i) Shown are the third-order nonlinear transverse voltage $V_{\beta\alpha\alpha\alpha}^{3\omega}$ as a function of the cubed bias current $I_{\rm \alpha}^3$ for the $'XZ'$ (red), $'YZ'$ (blue), and $'ZX'$ (grey) devices, respectively.}
    \label{fig:fig2}
\end{figure}

We have fabricated Hall bar devices to test the presence of the predicted third-order Hall response of altermagnetic CrSb to an external electric driving field. To this end, we have used focused ion beam (FIB) method to cut 1\,$\mu$m thin rectangular-shaped lamellae from a prism-shaped CrSb bulk crystal, which has a hexagonal cross section, as shown in Fig.~\ref{fig:fig2}(b). Following our symmetry analysis, lamellae were cut within the $\,xz-,\,yz-,\,\text{and}\,\,zx-$planes of the CrSb crystal to examine the anisotropic nature of the third-order Hall response. The resulting device samples are labeled $\,'XZ',\,'YZ',\,\text{and}\,\,'ZX'$. In the next step, the lamellae were placed on prepatterned gold electrodes, and electric contact between the lamellae and gold electrodes was established via FIB-deposited platinum (see Methods section for device fabrication details). Fig.~\ref{fig:fig2}(c) shows a scanning-electron microscopy (SEM) image of the $'XZ'$ device; SEM images of all other devices are shown in Sec.~III of the Suppl.~Materials. Note that we obtain $'XZ'$ and $'ZX'$ devices by rotating one of the crystalline lamellae by 90$^\circ$ before placing it on the electrodes.

We now focus on measurements of the nonlinear electric response of CrSb in a current-biased measurement geometry. We apply a sinusoidal a.c. excitation current of frequency $1\omega$ along the $\alpha$ direction and record the longitudinal $V_{\alpha\alpha\alpha\alpha}^{3\omega}$ and transverse $V_{\beta\alpha\alpha\alpha}^{3\omega}$ third-order voltage responses at the third harmonic $3\omega$ using lock-in measurements. The data and detailed analysis of the longitudinal third-response are presented in Sec.~IV of the Suppl.~Materials. In Fig.~\ref{fig:fig2}, (d)-(f) we schematically illustrate the specific measurement geometry of the $\,'XZ',\,'ZX',\,\text{and}\,\,'YZ'$ devices respectively. For example, when a bias current is applied along the $x$-direction of the $'XZ'$ cut in Fig.~\ref{fig:fig2}(d), the longitudinal and transverse third-order voltage responses correspond to $V_{xxxx}^{3\omega}$ and $V_{zxxx}^{3\omega}$, respectively. As can be seen, the choice of samples $\,'XZ',\,'YZ',\,\text{and}\,\,'ZX'$ enables us to examine the nonlinear electric response of CrSb when the a.c. current bias is applied along the $x-$, $y-$, and $z-$directions of the crystal. Control experiments that confirm the accuracy of our lock-in detection scheme are presented in Sec.~V of the Suppl.~Materials.

In Fig.~\ref{fig:fig2}(g)-(i), we display the transverse third-order voltage response $V_{\beta\alpha\alpha\alpha}^{3\omega}$ recorded on Hall bar devices $\,'XZ',\,'ZX',\,\text{and}\,\,'YZ'$ at room temperature as a function of the cubed applied bias current. Our measurements show that the $'XZ'$ device exhibits a distinct $V_{zxxx}^{3\omega}$ signal and $V_{zxxx}^{3\omega}$ scales cubically with the bias current (see linear fit as black dashed line). This establishes $V_{zxxx}^{3\omega}$ as a third-order NLAHE. Interestingly, the detection of the NLAHE is highly anisotropic. Measurements conducted on the $'ZX'$ and the $'YZ'$ devices do not detect signatures of a third-order Hall effect, as seen in Fig.~\ref{fig:fig2}, (h) and (i). The observation of the NLAHE in the $'XZ'$ device presents the key result of our work, and is reproduced on another device $'XZ2'$, as shown in Ext.~Data~Fig.~\ref{fig:extfig1}. The distinct spatial anisotropy of this measurement signal is in excellent agreement with results of our symmetry analysis, summarized in Fig.~\ref{fig:fig2}(a), suggesting that $V_{zxxx}^{3\omega}$ is induced by a Berry curvature quadrupole $\mathcal{Q}_{xxxz}$ within the spin-split band structure of CrSb.

We have also characterized the first and second order transverse voltage responses. Our measurements, presented in Sec.~VI of the Suppl.~Materials, show that none of the devices exhibits a longitudinal ($V_{\alpha\alpha\alpha}^{2\omega}$) or transverse ($V_{\beta\alpha\alpha}^{2\omega}$) second-order voltage response that scales quadratically with the applied bias current. This finding is consistent with results from our symmetry analysis and the general understanding that CrSb does not break $\mathcal{P}$. However, we detect a small transverse first-order voltage $V_{\beta\alpha}^{1\omega}$ in all devices. This observation is unexpected because our preceding analysis shows that the symmetries of CrSb prohibit the appearance of a Hall effect. Magnetic field $B$ dependent measurements, shown in Sec.~VI of the Suppl.~Materials, reveal that $V_{\beta\alpha}^{1\omega}$ is symmetric with respect to the reversal of the $B$-field direction, suggesting that this signal arises from anisotropy effects rather than being a true Hall signal. Indeed, an extended symmetry analysis, presented in Sec.~VII of the Suppl.~Materials, confirms that this first-order signal most likely arises from small angular misalignments of the sample plane with the principal crystallographic axes. Such small angular misalignment is a consequence of the alignment accuracy of about $0.5$° of the CrSb bulk crystal in the FIB cutting process (see Methods section).

\section{Experimental characterization of the third-order NLAHE in CrSb}

The results of our symmetry analysis suggest that the experimentally detected third-order NLAHE in device $'XZ'$ is induced by a BCQ that exists within the $xz-$plane of the Brillouin zone. At the same time, other mechanisms such as skew scattering~\cite{smit1955spontaneous, smit1958spontaneous}, a longitudinal third-order Drude response~\cite{zhang2023higher}, Joule heating~\cite{dames20051omega}, and an out-of-plane magnetization can also contribute or give rise to a third order transverse voltage. To strengthen the causality between the observed third-order NLAHE and a BCQ supported by the non-relativistic spin-split band structure, we have conducted further magnetic field and temperature dependent measurements. 

\begin{figure}[H]
    \centering
    \includegraphics[width=1\linewidth]{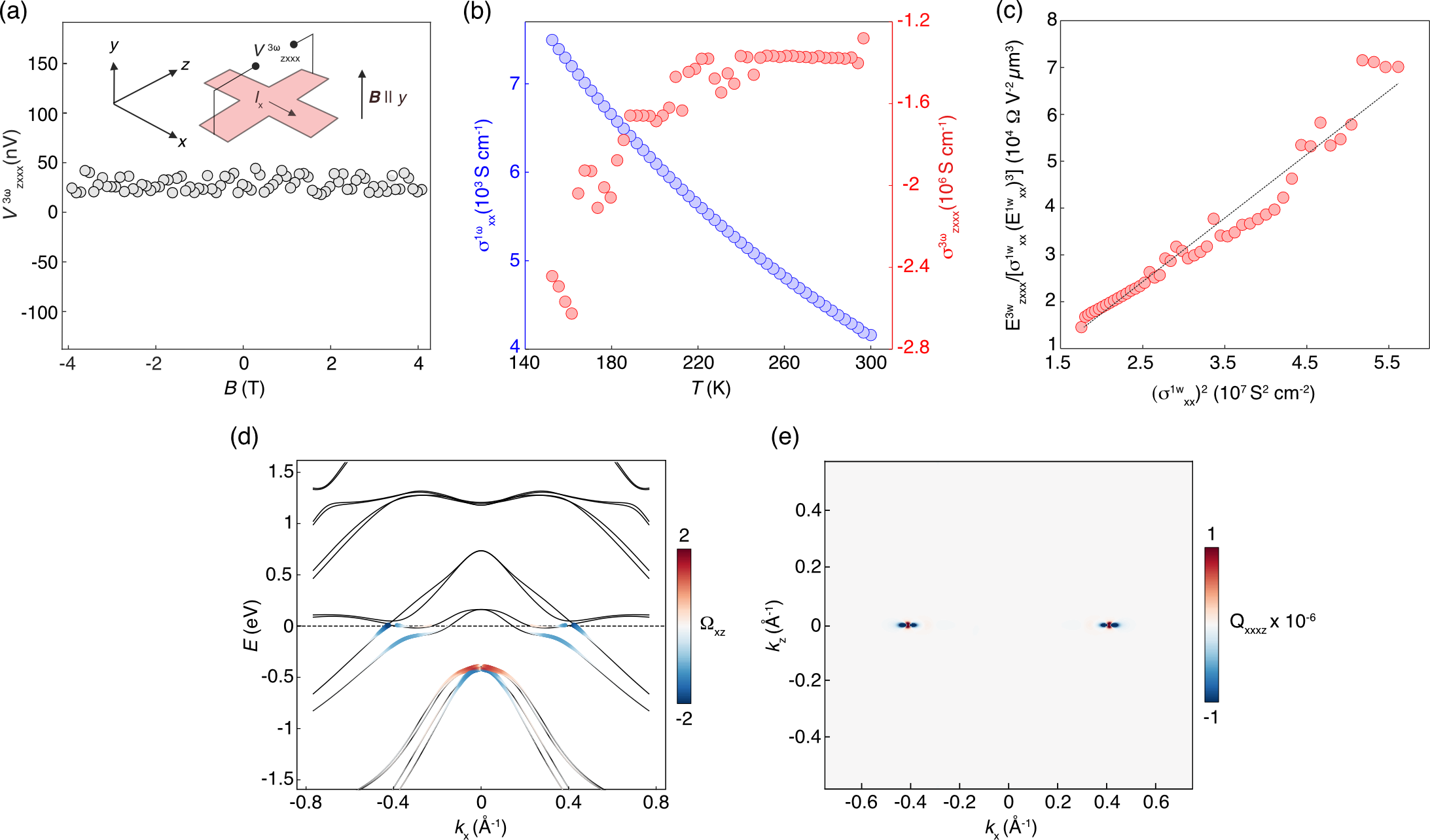}
    \caption{\textbf{Characterization of the third-order NLAHE of CrSb.} (a) Shown is the third-order nonlinear transverse voltage $V_{zxxx}^{3\omega}$ as a function of an external magnetic field (${\it B}$) applied along the $y-$direction. 
    (b) Shown are the first-order longitudinal conductivity $\sigma_{xx}^{1\omega}$ (blue symbols) and the third-order transverse conductivity $\sigma_{zxxx}^{3\omega}$ (red symbols) as a function of temperature $T$.
    (c) Shown is $E_{zxxx}^{3\omega}/(\sigma_{xx}^{1\omega}(E_{xx}^{1\omega})^3)$ and fit to the data (dashed line) plotted as a function of the squared longitudinal conductivity $\left( \sigma_{\rm xx}^{1\omega} \right)^2$.
    (d) Shown is the electronic band structure of CrSb, as obtained from DFT calculations (see Methods section), overlaid with the amplitude of the Berry curvature  $\Omega_{xz}$. The band structure is plotted along the $k_z$ direction at $k_y=0$ and $k_z=0.4\,Å^{-1}$.
    (e) Shown is the momentum space distribution of  $Q_{xxxz}$ within the $xz-$plane of the Brillouin zone at Fermi energy.}
    \label{fig:fig3}
\end{figure}

Like conventional antiferromagnets, altermagnets have a compensated magnetization ${\bf M }=0$. However, relativistic spin-orbit coupling could theoretically give rise to a small out-of-plane magnetization $M_z$ resulting in an anomalous contribution to the Hall effect~\cite{mazin2024origin}. Moreover, Joule heating could cause a longitudinal third-order voltage~\cite{dames20051omega}, which in the presence of a small net magnetization $M_z$ could give rise to a third-order NLAHE. The amplitude of these contributions can be generally tuned using an external magnetic field $B_z$ because $M_z\propto B_z$. Therefore, to exclude scenarios in which the third-order NLAHE would be caused by a net magnetization or Joule heating, we have studied the dependence of $V_{zxxx}^{3\omega}$ on an external magnetic field applied along the out-of-plane direction ($y$-axis for the $'XZ'$ device). As shown in Fig.~\ref{fig:fig3}(a), $V_{zxxx}^{3\omega}$ is finite at $B=0$ and remains unchanged for magnetic field sweeps between $-40\,$kOe and $40\,$kOe. This observation suggests that $V_{zxxx}^{3\omega}$ is not induced by a net magnetization or Joule heating that would couple to $\bf B$ but rather represents an intrinsic nonlinear material response to a longitudinal a.c. current drive.

Next, we examine the temperature ($T$)-evolution of the third-order NLAHE. In general, different mechanisms, such as skew scattering and BCQ, can contribute to the measurement signal $V_{zxxx}^{3\omega}$. Critically, each of these mechanisms is characterized by a distinct dependence on the charge carrier scattering time $\tau$, which is generally temperature dependent. Therefore, a scaling law analysis in terms of $\tau$ can be used to extract their respective contributions to $V_{zxxx}^{3\omega}$. To this end, we record the longitudinal first-order $\sigma_{xx}^{1\omega}(T)$ and third-order nonlinear anomalous Hall conductivity $\sigma_{zxxx}^{3\omega}(T)$ as a function of temperature $T$ between room temperature ($T=300\,$K) and 150\,K. As seen in Fig.~\ref{fig:fig3}(b), $\sigma_{xx}^{1\omega}(T)$ monotonically increases toward lower temperatures, consistent with the metallic character of CrSb. In contrast, $\sigma_{zxxx}^{3\omega}(T)$ reduces by about a factor of two over the same temperature range. Note that lowering the temperature even further resulted in contact degradation caused by thermal stress for all CrSb devices that we had fabricated, and the effect of this degradation was strong for $\sigma_{zxxx}^{3\omega}(T)$; this currently prevents us from examining $V_{zxxx}^{3\omega}$ over an even larger temperature range. 

To conduct a scaling law analysis, we plot $E_{zxxx}^{3\omega}/(\sigma_{xx}^{1\omega}(E_{xx}^{1\omega})^3)$ as a function of the squared longitudinal conductivity in Fig.~\ref{fig:fig3}(c). $E_{zxxx}^{3\omega}$ and $E_{xx}^{1\omega}$ denote the electric field amplitudes of $V_{zxxx}^{3\omega}$ and $V_{xx}^{1\omega}$, respectively. It was previously shown that BCQ contributions $\beta$ to the third-order transverse voltage can be determined using the scaling law $E_{zxxx}^{3\omega}/(\sigma_{xx}^{1\omega}(E_{xx}^{1\omega})^3)=\alpha\sigma_{xx}^2+\beta$, where $\alpha$ accounts for contributions from skew scattering and BCQ contributions can be identified and extracted from the vertical intercept~\cite{sankar2024experimental}. We can fit the data using this scaling law expression, as seen in Fig.~\ref{fig:fig3}(c). The finite vertical intercept $\beta=(-9.77\pm0.07)\times10^3\,\Omega\mu\text{m}^3V^{-2}$, extracted from the fit, supports a BCQ-induced third-order NLAHE that accounts for $\approx70\,\%$ of the detected $V_{zxxx}^{3\omega}$ amplitude at room temperature, while the linear term $\alpha=(1032\pm8)\times10^{-6}\Omega^3\mu\text{m}^5V^{-2}$ indicates the presence of skew-scattering contributions that become more dominant at lower temperatures. Therefore, the scaling law analysis of $V_{zxxx}^{3\omega}$ is consistent with the results of our symmetry analysis, suggesting the observation of a BCQ-induced third-order NLAHE. Note that a comparable analysis of the longitudinal third-order voltages, presented in Sec.~IV of the Suppl.~Materials, reveals that $V_{\alpha, \alpha,\alpha, \alpha}^{3\omega}$ is dominated by contributions from a QMQ and Drude scattering, which is consistent with our symmetry analysis presented in Fig.~\ref{fig:fig2}(a).

\section{Berry curvature quadrupole in the spin-split band structure of C\lowercase{r}S\lowercase{b}}

We further present results from density functional theory (DFT) calculations to validate that the electronic band structure of CrSb supports a BCQ within $xz-$plane of the Brillouin zone. Calculation details are presented in the Methods section. In the absence of relativistic SOC, spin remains a good quantum number, and spin-splitting in the electronic band structure of CrSb is induced by non-relativistic antiferromagnetic exchange energy. This process results in numerous band crossing points of opposite-spin polarization near the Fermi energy. Introducing SOC lifts these band degeneracy points, whose small spectral gaps act as Berry curvature hotspots near Fermi energy, as seen in Fig.~\ref{fig:fig3}(d). 

To understand the impact of these hotspots on the nonlinear third-order susceptibility of CrSb, we consider the projection of the electronic states into the relevant crystallographic direction. According to our magnetic point group analysis (see Sec.~II of the Suppl.~Mat.), the combined crystalline and magnetic symmetries support exactly one BCQ tensor element $\mathcal{Q}_{xxxz}$ within the $xz-$plane of the crystal, which would give rise to the observed third-order NLAHE $V_{zxxx}^{3\omega}$. To validate this understanding in terms of the realistic band structure, we plot the Fermi surface overlaid with the second derivative $\mathcal{Q}_{xxxz}(k_x,\,k_z)=\partial_x\partial_x\Omega_{xz}$ of the Berry curvature $\Omega_{xz}$ within the $xz$-plane of the Brillouin zone, as shown in Fig.~\ref{fig:fig3}(e). As can be seen, the band structure of CrSb supports a distinct BCQ distribution within the $xz-$plane, whose momentum-space characteristics are determined by the underlying crystalline and magnetic symmetries. An extended symmetry analysis of the BCQ distribution in the Brillouin zone is presented in Sec.~VIII of the Suppl. Materials. Critically, we also find that the momentum space integration of $\mathcal{Q}_{xxxz}(k_x,\,k_z)$ results in a finite contribution to $\sigma_{zxxx}^{3\omega}$ at Fermi energy, as seen in Ext.Data.Fig.~\ref{fig:extfig2}. Hence, consistent with results from our symmetry and scaling law analyses, our band structure analysis suggests that the spin-split electronic bands of CrSb support a BCQ-induced third-order NLAHE.

Note that experimentally determined and theoretical values of the linear anomalous Hall conductivity are often quantitatively compared to support the interpretation of the experimental observations~\cite{tian2009proper, nagaosa2010anomalous}. In the case of BCQ, we point out that such comparison is generally challenging. Owing to the $E_{zxxx}^{3\omega}/(\sigma_{xx}^{1\omega}(E_{xx}^{1\omega})^3)$ form of the scaling law~\cite{sankar2024experimental}, the extracted BCQ contribution to $\sigma_{zxxx}^{3\omega}$ is a function of $\sigma_{xx}^{1\omega}$, where $E_{xx}^{1\omega}\propto(\sigma_{xx}^{1\omega})^{-1}$. Hence, the extracted BCQ value is influenced by various material properties, such as charge carrier density and electron mobility, determined by the details of the electronic band structure, impurity concentration, and temperature, effects that are not included in the calculation of the BCQ using a Boltzmann equation approach based on a DFT-derived tight-binding model, making a quantitative comparison very challenging.

\section{Room temperature multiple wave mixing in an altermagnet}

\begin{figure}[H]
    \centering
    \includegraphics[width=1\linewidth]{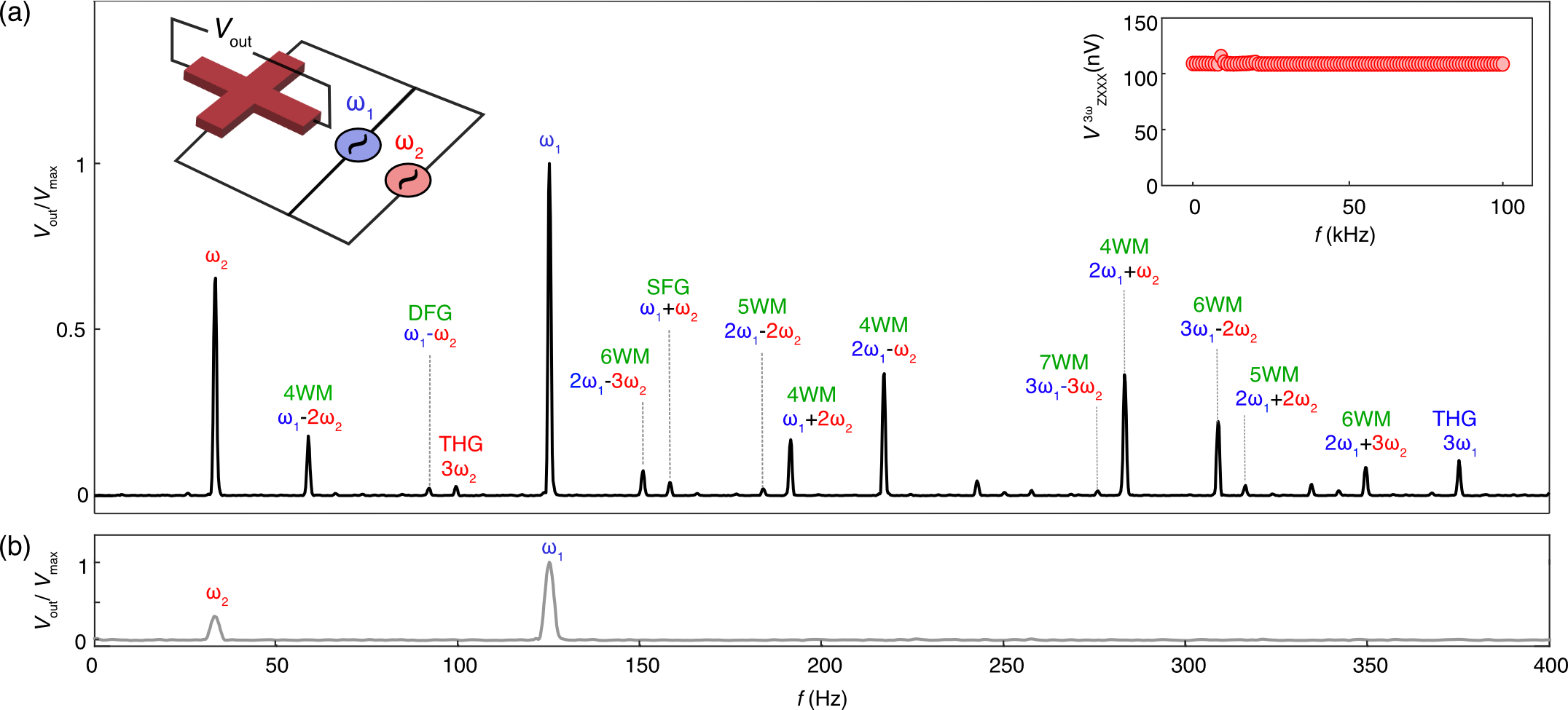}
    \caption{\textbf{Broadband multiple wave mixing in altermagnetic CrSb.} (a) Shown is the frequency ($f$) spectrum of the output voltage $V_{\rm out}$ of a multiple wave mixer based on the $'XZ'$ device. $V_{\rm out}$ is normalized to the maximum output amplitude $V_{\rm max}$ seen at $\omega_1$. As schematically illustrated in the left inset two a.c. currents of different frequencies ($\omega_1 = 125\,$Hz, $\omega_2 = 33\,$Hz) were simultaneously applied to the 'XZ' Hall bar device and the frequency-dependent Hall voltage $V_{\rm out}$ was recorded. The detected output signals at different frequencies are labeled according to their origin, as described in the main text. The right inset displays the third-order nonlinear transverse voltage $V_{zxxx}^{3\omega}$ as a function of frequency $f =0-100$ kHz of the longitudinal input current $I_x$. (b) Shown is the result of a control experiment where the frequency-dependent output voltage $V_{\rm out}$ was recorded when $\omega_1$ and $\omega_2$ are applied to a simple resistor with electric resistance $R=100\,\Omega$. $V_{\rm out}$ is normalized to the maximum output amplitude $V_{\rm max}$ seen at $\omega_1$.}
    \label{fig:fig4}
\end{figure}

Our combined experimental and theoretical study provides evidence for a BCQ induced third-order NLAHE in altermagnetic CrSb. In general, NLAHEs hold promise for technological applications in the field of high-frequency signal generation, conversion, and rectification, as well as energy harvesting of ambient RF signals~\cite{du2021nonlinear}. To quantify the potential of altermagnets in this context, we first consider the third-order nonlinear susceptibility $\chi_{xxxz}^{3\omega}$ of the BCQ. It can be derived from the Boltzmann equation and is typically of the form
\begin{equation}
    \chi_{xxxz}^{3\omega}\propto\frac{Q_{xxxz}}{(1-i\omega\tau)(1-2i\omega\tau)(1-3i\omega\tau)}
\end{equation} 
and captures the inertia of the BCQ in response to an external driving field. It follows that the bandwidth of the third-order nonlinear Hall response is limited by the charge carrier relaxation time $\tau=\mu\,m_{\rm eff}/e$, where $\mu$, $m_{\rm eff}$, and $e$ denote the charge carrier mobility, effective mass, and electron charge, respectively. Using experimentally determined values of $\mu$ and $m_{\rm eff}$ for CrSb~\cite{urata2024high}, we estimate $\tau\approx1\,$ps. This time scale corresponds to the maximum bandwidth $\delta f=1/\tau\approx1\,$THz of $\chi_{xxxz}^{3\omega}$. To experimentally test the bandwidth of the NLAHE in CrSb, we have recorded the amplitude of $V_{zxxx}^{3\omega}$ as a function of the frequency $\omega$ of the longitudinal a.c. drive current $I_{\rm x}$. As can be seen in the inset of Fig.~\ref{fig:fig4}(a), $V_{zxxx}^{3\omega}$ remains constant over the entire experimentally accessible frequency range $f\leq100\,$kHz. These broadband characteristics of the NLAHE in CrSb are consistent with our above estimate for $\chi_{xxxz}^{3\omega}$.

Finally, we utilize this nonlinear electric response of altermagnetic CrSb to realize a broadband multiple wave mixing (MWM) device. MWM can occur in materials with a third-order nonlinear susceptibility $\chi^{3\omega}$ and describes the combination of several input electromagnetic waves into output waves of different frequency. This phenomenon is widely used to realize coherent tunable light sources in parametric oscillators, in quantum-limited amplification, and communication networks and high-resolution spectroscopy. While MWM is typically realized using the nonlinearity of atoms and superconducting circuits~\cite{maker1965study, carman1966observation, roch2012widely}, more recent studies also discuss this phenomenon in the context of nonlinear Hall effects~\cite{min2024colossal, kiyonaga2025ac}. Here, we show that MWM can be realized in altermagnets using the $\chi_{xxxz}^{3\omega}$ of the BCQ in CrSb. 

We implement a MWM device by connecting two current sources $I_{\rm x}^{1\omega_1}$ and $I_{\rm x}^{1\omega_2}$ with identical current amplitude but different frequencies $\omega_1$ and $\omega_2$ in parallel along the longitudinal direction of device $'XZ'$, as shown in the inset of Fig.~\ref{fig:fig4}. We then record the frequency spectrum of $V_{zxxx}(f)$. When two alternating currents with $\omega_1=125\,$Hz and $\omega_2=33\,$Hz are applied, our measurements show that $V_{zxxx}(f)$ exhibits about 20 pronounced spectral peaks between 0 to 500\,Hz, as seen in Fig.~\ref{fig:fig4}(a). In addition to the fundamental ($\omega_1$ and $\omega_2$) and third harmonic ($3\omega_1$ and $3\omega_2$) frequencies of the input signals, we also detect spectral components that result from sum frequency generation (SFG, $\omega_1+\omega_2$) and different frequency generation (DFG, $\omega_1-\omega_2$) as well as MWM of the input signals. The MWM output appears at frequencies $\omega=\alpha\omega_1\pm\beta\omega_2$ where $\alpha$ and $\beta$ are positive integers. We find that the four wave mixing (4WM) signals $V_{\rm zxxx}^{2\omega_1\pm\omega_2}$ at $\omega=2\omega_1-\omega_2=217\,$Hz and $\omega=2\omega_1+\omega_2=283\,$Hz exhibit the largest output amplitude next to the fundamental harmonics $\omega_1$ and $\omega_2$. Critically, the occurrence of 4WM in a nonlinear medium requires a finite third-order nonlinear susceptibility, which here is provided by $\chi_{xxxz}^{3\omega}$ of the Berry curvature quadrupole. We have also conducted a control experiment by replacing the $'XZ'$ device with a simple resistor of electric resistance $R=100\,\Omega$. Repeating the same measurement as before, we find that the transverse output signal does not exhibit SFG, DFG, and MWM components, as seen in Fig.~\ref{fig:fig4}(b). Only small spectral peaks can be detected at the fundamental frequencies $\omega_1$ and $\omega_2$, which arise from inductive coupling in the measurement circuit. These results establish the realization of a broadband mutliple wave mixing device based on the intrinsic nonlinear susceptibility of CrSb.

To conclude, our discovery of a Berry curvature quadrupole induced NLAHE in the paradigmatic altermagnet CrSb experimentally establishes that the intrinsic electric response of fully compensated altermagnets~\cite{fang2024quantum, ezawa2024intrinsic} is governed by the underlying magnetic crystal order~\cite{vsmejkal2022anomalous, hu2025catalog}. Hence, our study provides a generalized understanding of the impact of the non-relativistic spin splitting band structure on the electric material properties of altermagnets beyond the linear response regime previously studied in the limit of relativistic spin-orbit coupling and finite magnetization~\cite{feng2022anomalous, gonzalez2023spontaneous, kluczyk2024coexistence, chilcote2024stoichiometry, mazin2024origin}. Existing concepts for the potential use of altermagnets in spintronics applications are confined to $d$-wave altermagnets owing to symmetry-constraints~\cite{bose2022tilted, bai2022observation, karube2022observation, gonzalez2021efficient, bai2023efficient, shao2023neel, zhou2024crystal, badura2025observation}. Through the realization of broadband multiple wave mixing device, which utilizes the nonlinear susceptibility $\chi^{3\omega}$ of the Berry curvature quadrupole embedded in the spin-split electronic structure of CrSb, our study overcomes these limitations and showcases the general potential of $(d,\,g,\,i)-$wave altermagnets in applications for high-frequency electronics, THz generation, communication networks, and energy harvesting.

\clearpage
\bibliography{bibliography}

@article{hayami2020bottom,
  title={Bottom-up design of spin-split and reshaped electronic band structures in antiferromagnets without spin-orbit coupling: Procedure on the basis of augmented multipoles},
  author={Hayami, Satoru and Yanagi, Yuki and Kusunose, Hiroaki},
  journal={Physical Review B},
  volume={102},
  number={14},
  pages={144441},
  year={2020},
  publisher={APS}
}

@article{ma2021multifunctional,
  title={Multifunctional antiferromagnetic materials with giant piezomagnetism and noncollinear spin current},
  author={Ma, Hai-Yang and Hu, Mengli and Li, Nana and Liu, Jianpeng and Yao, Wang and Jia, Jin-Feng and Liu, Junwei},
  journal={Nature communications},
  volume={12},
  number={1},
  pages={2846},
  year={2021},
  publisher={Nature Publishing Group UK London}
}

@article{vsmejkal2022beyond,
  title={Beyond conventional ferromagnetism and antiferromagnetism: A phase with nonrelativistic spin and crystal rotation symmetry},
  author={{\v{S}}mejkal, Libor and Sinova, Jairo and Jungwirth, Tomas},
  journal={Physical Review X},
  volume={12},
  number={3},
  pages={031042},
  year={2022},
  publisher={APS}
}

@article{krempasky2024altermagnetic,
  title={Altermagnetic lifting of Kramers spin degeneracy},
  author={Krempask{\`y}, J and {\v{S}}mejkal, L and D’souza, SW and Hajlaoui, M and Springholz, G and Uhl{\'\i}{\v{r}}ov{\'a}, K and Alarab, F and Constantinou, PC and Strocov, V and Usanov, D and others},
  journal={Nature},
  volume={626},
  number={7999},
  pages={517--522},
  year={2024},
  publisher={Nature Publishing Group UK London}
}

@article{reimers2024direct,
  title={Direct observation of altermagnetic band splitting in CrSb thin films},
  author={Reimers, Sonka and Odenbreit, Lukas and {\v{S}}mejkal, Libor and Strocov, Vladimir N and Constantinou, Procopios and Hellenes, Anna B and Jaeschke Ubiergo, Rodrigo and Campos, Warlley H and Bharadwaj, Venkata K and Chakraborty, Atasi and others},
  journal={Nature Communications},
  volume={15},
  number={1},
  pages={2116},
  year={2024},
  publisher={Nature Publishing Group UK London}
}

@article{ding2024large,
  title={Large band splitting in g-wave altermagnet CrSb},
  author={Ding, Jianyang and Jiang, Zhicheng and Chen, Xiuhua and Tao, Zicheng and Liu, Zhengtai and Li, Tongrui and Liu, Jishan and Sun, Jianping and Cheng, Jinguang and Liu, Jiayu and others},
  journal={Physical Review Letters},
  volume={133},
  number={20},
  pages={206401},
  year={2024},
  publisher={APS}
}

@article{zhang2025crystal,
  title={Crystal-symmetry-paired spin--valley locking in a layered room-temperature metallic altermagnet candidate},
  author={Zhang, Fayuan and Cheng, Xingkai and Yin, Zhouyi and Liu, Changchao and Deng, Liwei and Qiao, Yuxi and Shi, Zheng and Zhang, Shuxuan and Lin, Junhao and Liu, Zhengtai and others},
  journal={Nature Physics},
  pages={1--8},
  year={2025},
  publisher={Nature Publishing Group UK London}
}

@article{jiang2025metallic,
  title={A metallic room-temperature d-wave altermagnet},
  author={Jiang, Bei and Hu, Mingzhe and Bai, Jianli and Song, Ziyin and Mu, Chao and Qu, Gexing and Li, Wan and Zhu, Wenliang and Pi, Hanqi and Wei, Zhongxu and others},
  journal={Nature Physics},
  pages={1--6},
  year={2025},
  publisher={Nature Publishing Group UK London}
}

@article{vsmejkal2022anomalous,
  title={Anomalous hall antiferromagnets},
  author={{\v{S}}mejkal, Libor and MacDonald, Allan H and Sinova, Jairo and Nakatsuji, Satoru and Jungwirth, Tomas},
  journal={Nature Reviews Materials},
  volume={7},
  number={6},
  pages={482--496},
  year={2022},
  publisher={Nature Publishing Group UK London}
}

@article{hu2025catalog,
  title={Catalog of C-Paired Spin-Momentum Locking in Antiferromagnetic Systems},
  author={Hu, Mengli and Cheng, Xingkai and Huang, Zhenqiao and Liu, Junwei},
  journal={Physical Review X},
  volume={15},
  number={2},
  pages={021083},
  year={2025},
  publisher={APS}
}

@article{urata2024high,
  title={High mobility charge transport in a multicarrier altermagnet CrSb},
  author={Urata, Takahiro and Hattori, Wataru and Ikuta, Hiroshi},
  journal={Physical Review Materials},
  volume={8},
  number={8},
  pages={084412},
  year={2024},
  publisher={APS}
}

@article{mazin2024origin,
  title={Origin of the gossamer ferromagnetism in MnTe},
  author={Mazin, Igor I and Belashchenko, KD},
  journal={Physical Review B},
  volume={110},
  number={21},
  pages={214436},
  year={2024},
  publisher={APS}
}

@article{gonzalez2023spontaneous,
  title={Spontaneous anomalous Hall effect arising from an unconventional compensated magnetic phase in a semiconductor},
  author={Gonzalez Betancourt, RD and Zub{\'a}{\v{c}}, Jan and Gonzalez-Hernandez, R and Geishendorf, Kevin and {\v{S}}ob{\'a}{\v{n}}, Zbynek and Springholz, Gunther and Olejn{\'\i}k, Kamil and {\v{S}}mejkal, Libor and Sinova, Jairo and Jungwirth, Tomas and others},
  journal={Physical Review Letters},
  volume={130},
  number={3},
  pages={036702},
  year={2023},
  publisher={APS}
}

@article{kluczyk2024coexistence,
  title={Coexistence of anomalous Hall effect and weak magnetization in a nominally collinear antiferromagnet MnTe},
  author={Kluczyk, KP and Gas, K and Grzybowski, MJ and Skupi{\'n}ski, P and Borysiewicz, MA and F{\k{a}}s, T and Suffczy{\'n}ski, J and Domagala, JZ and Grasza, K and Mycielski, A and others},
  journal={Physical Review B},
  volume={110},
  number={15},
  pages={155201},
  year={2024},
  publisher={APS}
}

@article{chilcote2024stoichiometry,
  title={Stoichiometry-induced ferromagnetism in altermagnetic candidate MnTe},
  author={Chilcote, Michael and Mazza, Alessandro R and Lu, Qiangsheng and Gray, Isaiah and Tian, Qi and Deng, Qinwen and Moseley, Duncan and Chen, An-Hsi and Lapano, Jason and Gardner, Jason S and others},
  journal={Advanced Functional Materials},
  volume={34},
  number={46},
  pages={2405829},
  year={2024},
  publisher={Wiley Online Library}
}

@article{feng2022anomalous,
  title={An anomalous Hall effect in altermagnetic ruthenium dioxide},
  author={Feng, Zexin and Zhou, Xiaorong and {\v{S}}mejkal, Libor and Wu, Lei and Zhu, Zengwei and Guo, Huixin and Gonz{\'a}lez-Hern{\'a}ndez, Rafael and Wang, Xiaoning and Yan, Han and Qin, Peixin and others},
  journal={Nature Electronics},
  volume={5},
  number={11},
  pages={735--743},
  year={2022},
  publisher={Nature Publishing Group UK London}
}

@article{sankar2024experimental,
  title={Experimental evidence for a Berry curvature quadrupole in an antiferromagnet},
  author={Sankar, Soumya and Liu, Ruizi and Zhang, Cheng-Ping and Li, Qi-Fang and Chen, Caiyun and Gao, Xue-Jian and Zheng, Jiangchang and Lin, Yi-Hsin and Qian, Kun and Yu, Ruo-Peng and others},
  journal={Physical Review X},
  volume={14},
  number={2},
  pages={021046},
  year={2024},
  publisher={APS}
}

@article{nagaosa2010anomalous,
  title={Anomalous hall effect},
  author={Nagaosa, Naoto and Sinova, Jairo and Onoda, Shigeki and MacDonald, Allan H and Ong, Nai Phuan},
  journal={Reviews of modern physics},
  volume={82},
  number={2},
  pages={1539--1592},
  year={2010},
  publisher={APS}
}

@article{smit1955spontaneous,
  title={The spontaneous Hall effect in ferromagnetics I},
  author={Smit, J},
  journal={Physica},
  volume={21},
  number={6-10},
  pages={877--887},
  year={1955},
  publisher={Elsevier}
}

@article{smit1958spontaneous,
  title={The spontaneous Hall effect in ferromagnetics II},
  author={Smit, Jan},
  journal={Physica},
  volume={24},
  number={1-5},
  pages={39--51},
  year={1958},
  publisher={Elsevier}
}

@article{dames20051omega,
  title={1$\omega$, 2$\omega$, and 3$\omega$ methods for measurements of thermal properties},
  author={Dames, Chris and Chen, Gang},
  journal={Review of scientific Instruments},
  volume={76},
  number={12},
  year={2005},
  publisher={AIP Publishing}
}

@article{tian2009proper,
  title={Proper scaling of the anomalous Hall effect},
  author={Tian, Yuan and Ye, Li and Jin, Xiaofeng},
  journal={Physical review letters},
  volume={103},
  number={8},
  pages={087206},
  year={2009},
  publisher={APS}
}

@article{zhang2023higher,
  title={Higher-order nonlinear anomalous Hall effects induced by Berry curvature multipoles},
  author={Zhang, Cheng-Ping and Gao, Xue-Jian and Xie, Ying-Ming and Po, Hoi Chun and Law, Kam Tuen},
  journal={Physical Review B},
  volume={107},
  number={11},
  pages={115142},
  year={2023},
  publisher={APS}
}

@article{min2024colossal,
  title={Colossal room-temperature non-reciprocal Hall effect},
  author={Min, Lujin and Zhang, Yang and Xie, Zhijian and Ayyagari, Sai Venkata Gayathri and Miao, Leixin and Onishi, Yugo and Lee, Seng Huat and Wang, Yu and Alem, Nasim and Fu, Liang and others},
  journal={Nature Materials},
  volume={23},
  number={12},
  pages={1671--1677},
  year={2024},
  publisher={Nature Publishing Group UK London}
}

@article{maker1965study,
  title={Study of optical effects due to an induced polarization third order in the electric field strength},
  author={Maker, PD and Terhune, RW},
  journal={Physical Review},
  volume={137},
  number={3A},
  pages={A801},
  year={1965},
  publisher={APS}
}

@article{carman1966observation,
  title={Observation of degenerate stimulated four-photon interaction and four-wave parametric amplification},
  author={Carman, RL and Chiao, RY and Kelley, PL},
  journal={Physical Review Letters},
  volume={17},
  number={26},
  pages={1281},
  year={1966},
  publisher={APS}
}

@article{roch2012widely,
  title={Widely Tunable, Nondegenerate Three-Wave Mixing Microwave Device Operating<? format?> near the Quantum Limit},
  author={Roch, Nicolas and Flurin, Emmanuel and Nguyen, Francois and Morfin, Pascal and Campagne-Ibarcq, Philippe and Devoret, Michel H and Huard, Benjamin},
  journal={Physical review letters},
  volume={108},
  number={14},
  pages={147701},
  year={2012},
  publisher={APS}
}

@article{takei1963magnetic,
  title={Magnetic structures in the MnSb-CrSb system},
  author={Takei, WJ and Cox, De E and Shirane, G},
  journal={Physical Review},
  volume={129},
  number={5},
  pages={2008},
  year={1963},
  publisher={APS}
}

@article{brinkman1966theory,
  title={Theory of spin-space groups},
  author={Brinkman, WF and Elliott, Roger James},
  journal={Proceedings of the Royal Society of London. Series A. Mathematical and Physical Sciences},
  volume={294},
  number={1438},
  pages={343--358},
  year={1966},
  publisher={The Royal Society London}
}

@article{litvin1974spin,
  title={Spin groups},
  author={Litvin, Daniel B and Opechowski, W},
  journal={Physica},
  volume={76},
  number={3},
  pages={538--554},
  year={1974},
  publisher={Elsevier}
}

@article{chen2024enumeration,
  title={Enumeration and representation theory of spin space groups},
  author={Chen, Xiaobing and Ren, Jun and Zhu, Yanzhou and Yu, Yutong and Zhang, Ao and Liu, Pengfei and Li, Jiayu and Liu, Yuntian and Li, Caiheng and Liu, Qihang},
  journal={Physical Review X},
  volume={14},
  number={3},
  pages={031038},
  year={2024},
  publisher={APS}
}

@article{jiang2024enumeration,
  title={Enumeration of spin-space groups: Toward a complete description of symmetries of magnetic orders},
  author={Jiang, Yi and Song, Ziyin and Zhu, Tiannian and Fang, Zhong and Weng, Hongming and Liu, Zheng-Xin and Yang, Jian and Fang, Chen},
  journal={Physical Review X},
  volume={14},
  number={3},
  pages={031039},
  year={2024},
  publisher={APS}
}

@article{snow1952neutron,
  title={Neutron diffraction investigation of the atomic magnetic moment orientation in the antiferromagnetic compound CrSb},
  author={Snow, AI},
  journal={Physical Review},
  volume={85},
  number={2},
  pages={365},
  year={1952},
  publisher={APS}
}

@article{abe1984magnetic,
  title={Magnetic properties of CrSb},
  author={Abe, Shunya and Kaneko, Takejiro and Ohashi, Masayoshi and Yoshida, Hajime and Kamigaki, Kazuo},
  journal={Journal of the Physical Society of Japan},
  volume={53},
  number={8},
  pages={2703--2709},
  year={1984},
  publisher={The Physical Society of Japan}
}

@article{yang2025three,
  title={Three-dimensional mapping of the altermagnetic spin splitting in CrSb},
  author={Yang, Guowei and Li, Zhanghuan and Yang, Sai and Li, Jiyuan and Zheng, Hao and Zhu, Weifan and Pan, Ze and Xu, Yifu and Cao, Saizheng and Zhao, Wenxuan and others},
  journal={Nature Communications},
  volume={16},
  number={1},
  pages={1442},
  year={2025},
  publisher={Nature Publishing Group UK London}
}

@article{lu2025signature,
  title={Signature of Topological Surface Bands in Altermagnetic Weyl Semimetal CrSb},
  author={Lu, Wenlong and Feng, Shiyu and Wang, Yuzhi and Chen, Dong and Lin, Zihan and Liang, Xin and Liu, Siyuan and Feng, Wanxiang and Yamagami, Kohei and Liu, Junwei and others},
  journal={Nano Letters},
  volume={25},
  number={18},
  pages={7343--7350},
  year={2025},
  publisher={ACS Publications}
}

@article{ma2019observation,
  title={Observation of the nonlinear Hall effect under time-reversal-symmetric conditions},
  author={Ma, Qiong and Xu, Su-Yang and Shen, Huitao and MacNeill, David and Fatemi, Valla and Chang, Tay-Rong and Mier Valdivia, Andr{\'e}s M and Wu, Sanfeng and Du, Zongzheng and Hsu, Chuang-Han and others},
  journal={Nature},
  volume={565},
  number={7739},
  pages={337--342},
  year={2019},
  publisher={Nature Publishing Group UK London}
}

@article{sodemann2015quantum,
  title={Quantum nonlinear Hall effect induced by Berry curvature dipole in time-reversal invariant materials},
  author={Sodemann, Inti and Fu, Liang},
  journal={Physical review letters},
  volume={115},
  number={21},
  pages={216806},
  year={2015},
  publisher={APS}
}

@article{gao2014field,
  title={Field induced positional shift of Bloch electrons and its dynamical implications},
  author={Gao, Yang and Yang, Shengyuan A and Niu, Qian},
  journal={Physical review letters},
  volume={112},
  number={16},
  pages={166601},
  year={2014},
  publisher={APS}
}

@article{lai2021third,
  title={Third-order nonlinear Hall effect induced by the Berry-connection polarizability tensor},
  author={Lai, Shen and Liu, Huiying and Zhang, Zhaowei and Zhao, Jianzhou and Feng, Xiaolong and Wang, Naizhou and Tang, Chaolong and Liu, Yuanda and Novoselov, KS and Yang, Shengyuan A and others},
  journal={Nature Nanotechnology},
  volume={16},
  number={8},
  pages={869--873},
  year={2021},
  publisher={Nature Publishing Group UK London}
}

@article{gao2023quantum,
  title={Quantum metric nonlinear Hall effect in a topological antiferromagnetic heterostructure},
  author={Gao, Anyuan and Liu, Yu-Fei and Qiu, Jian-Xiang and Ghosh, Barun and V. Trevisan, Tha{\'\i}s and Onishi, Yugo and Hu, Chaowei and Qian, Tiema and Tien, Hung-Ju and Chen, Shao-Wen and others},
  journal={Science},
  volume={381},
  number={6654},
  pages={181--186},
  year={2023},
  publisher={American Association for the Advancement of Science}
}

@article{wang2023quantum,
  title={Quantum-metric-induced nonlinear transport in a topological antiferromagnet},
  author={Wang, Naizhou and Kaplan, Daniel and Zhang, Zhaowei and Holder, Tobias and Cao, Ning and Wang, Aifeng and Zhou, Xiaoyuan and Zhou, Feifei and Jiang, Zhengzhi and Zhang, Chusheng and others},
  journal={Nature},
  volume={621},
  number={7979},
  pages={487--492},
  year={2023},
  publisher={Nature Publishing Group UK London}
}

@article{fang2024quantum,
  title={Quantum geometry induced nonlinear transport in altermagnets},
  author={Fang, Yuan and Cano, Jennifer and Ghorashi, Sayed Ali Akbar},
  journal={Physical Review Letters},
  volume={133},
  number={10},
  pages={106701},
  year={2024},
  publisher={APS}
}

@article{du2021nonlinear,
  title={Nonlinear hall effects},
  author={Du, ZZ and Lu, Hai-Zhou and Xie, XC},
  journal={Nature Reviews Physics},
  volume={3},
  number={11},
  pages={744--752},
  year={2021},
  publisher={Nature Publishing Group UK London}
}

@article{mostofi2008wannier90,
  title={wannier90: A tool for obtaining maximally-localised Wannier functions},
  author={Mostofi, Arash A and Yates, Jonathan R and Lee, Young-Su and Souza, Ivo and Vanderbilt, David and Marzari, Nicola},
  journal={Computer physics communications},
  volume={178},
  number={9},
  pages={685--699},
  year={2008},
  publisher={Elsevier}
}

@article{kresse1996efficient,
  title={Efficient iterative schemes for ab initio total-energy calculations using a plane-wave basis set},
  author={Kresse, Georg and Furthm{\"u}ller, J{\"u}rgen},
  journal={Physical review B},
  volume={54},
  number={16},
  pages={11169},
  year={1996},
  publisher={APS}
}

@article{kresse1996efficiency,
  title={Efficiency of ab-initio total energy calculations for metals and semiconductors using a plane-wave basis set},
  author={Kresse, Georg and Furthm{\"u}ller, J{\"u}rgen},
  journal={Computational materials science},
  volume={6},
  number={1},
  pages={15--50},
  year={1996},
  publisher={Elsevier}
}

@article{perdew1996generalized,
  title={Generalized gradient approximation made simple},
  author={Perdew, John P and Burke, Kieron and Ernzerhof, Matthias},
  journal={Physical review letters},
  volume={77},
  number={18},
  pages={3865},
  year={1996},
  publisher={APS}
}

@article{ezawa2024intrinsic,
  title={Intrinsic nonlinear conductivity induced by quantum geometry in altermagnets and measurement of the in-plane N{\'e}el vector},
  author={Ezawa, Motohiko},
  journal={Physical Review B},
  volume={110},
  number={24},
  pages={L241405},
  year={2024},
  publisher={APS}
}

@article{bose2022tilted,
  title={Tilted spin current generated by the collinear antiferromagnet ruthenium dioxide},
  author={Bose, Arnab and Schreiber, Nathaniel J and Jain, Rakshit and Shao, Ding-Fu and Nair, Hari P and Sun, Jiaxin and Zhang, Xiyue S and Muller, David A and Tsymbal, Evgeny Y and Schlom, Darrell G and others},
  journal={Nature Electronics},
  volume={5},
  number={5},
  pages={267--274},
  year={2022},
  publisher={Nature Publishing Group UK London}
}

@article{bai2022observation,
  title={Observation of spin splitting torque in a collinear antiferromagnet RuO 2},
  author={Bai, Hua and Han, Lei and Feng, XY and Zhou, YJ and Su, RX and Wang, Qian and Liao, LY and Zhu, WX and Chen, XZ and Pan, Feng and others},
  journal={Physical Review Letters},
  volume={128},
  number={19},
  pages={197202},
  year={2022},
  publisher={APS}
}

@article{karube2022observation,
  title={Observation of spin-splitter torque in collinear antiferromagnetic RuO 2},
  author={Karube, Shutaro and Tanaka, Takahiro and Sugawara, Daichi and Kadoguchi, Naohiro and Kohda, Makoto and Nitta, Junsaku},
  journal={Physical review letters},
  volume={129},
  number={13},
  pages={137201},
  year={2022},
  publisher={APS}
}

@article{gonzalez2021efficient,
  title={Efficient electrical spin splitter based on nonrelativistic collinear antiferromagnetism},
  author={Gonz{\'a}lez-Hern{\'a}ndez, Rafael and {\v{S}}mejkal, Libor and V{\`y}born{\`y}, Karel and Yahagi, Yuta and Sinova, Jairo and Jungwirth, Tom{\'a}{\v{s}} and {\v{Z}}elezn{\`y}, Jakub},
  journal={Physical Review Letters},
  volume={126},
  number={12},
  pages={127701},
  year={2021},
  publisher={APS}
}

@article{bai2023efficient,
  title={Efficient spin-to-charge conversion via altermagnetic spin splitting effect in antiferromagnet RuO 2},
  author={Bai, H and Zhang, YC and Zhou, YJ and Chen, P and Wan, CH and Han, L and Zhu, WX and Liang, SX and Su, YC and Han, XF and others},
  journal={Physical review letters},
  volume={130},
  number={21},
  pages={216701},
  year={2023},
  publisher={APS}
}

@article{shao2023neel,
  title={N{\'e}el spin currents in antiferromagnets},
  author={Shao, Ding-Fu and Jiang, Yuan-Yuan and Ding, Jun and Zhang, Shu-Hui and Wang, Zi-An and Xiao, Rui-Chun and Gurung, Gautam and Lu, WJ and Sun, YP and Tsymbal, Evgeny Y},
  journal={Physical Review Letters},
  volume={130},
  number={21},
  pages={216702},
  year={2023},
  publisher={APS}
}

@article{zhou2024crystal,
  title={Crystal thermal transport in altermagnetic RuO 2},
  author={Zhou, Xiaodong and Feng, Wanxiang and Zhang, Run-Wu and {\v{S}}mejkal, Libor and Sinova, Jairo and Mokrousov, Yuriy and Yao, Yugui},
  journal={Physical review letters},
  volume={132},
  number={5},
  pages={056701},
  year={2024},
  publisher={APS}
}

@article{badura2025observation,
  title={Observation of the anomalous Nernst effect in altermagnetic candidate Mn5Si3},
  author={Badura, Anton{\'\i}n and Campos, Warlley H and Bharadwaj, Venkata K and Kounta, Isma{\"\i}la and Michez, Lisa and Petit, Matthieu and Rial, Javier and Leivisk{\"a}, Miina and Baltz, Vincent and Krizek, Filip and others},
  journal={Nature Communications},
  volume={16},
  number={1},
  pages={7111},
  year={2025},
  publisher={Nature Publishing Group UK London}
}

@article{kiyonaga2025ac,
  title={AC Current-Driven Magnetization Switching and Nonlinear Hall Rectification in a Magnetic Topological Insulator},
  author={Kiyonaga, Yuto and Mogi, Masataka and Yoshimi, Ryutaro and Fujishiro, Yukako and Suzuki, Yuri and Birch, Max T and Tsukazaki, Atsushi and Kawamura, Minoru and Kawasaki, Masashi and Tokura, Yoshinori},
  journal={Advanced Materials},
  pages={e06210},
  year={2025},
  publisher={Wiley Online Library}
}

@Article{Zhou2025Manipulation,
author={Zhou, Zhiyuan
and Cheng, Xingkai
and Hu, Mengli
and Chu, Ruiyue
and Bai, Hua
and Han, Lei
and Liu, Junwei
and Pan, Feng
and Song, Cheng},
title={Manipulation of the altermagnetic order in CrSb via crystal symmetry},
journal={Nature},
year={2025},
month={Feb},
day={01},
volume={638},
number={8051},
pages={645-650},
issn={1476-4687},
doi={10.1038/s41586-024-08436-3},
url={https://doi.org/10.1038/s41586-024-08436-3}
}

@article{xiao2024spin,
  title={Spin space groups: Full classification and applications},
  author={Xiao, Zhenyu and Zhao, Jianzhou and Li, Yanqi and Shindou, Ryuichi and Song, Zhi-Da},
  journal={Physical Review X},
  volume={14},
  number={3},
  pages={031037},
  year={2024},
  publisher={APS}
}

@article{wu2007fermi,
  title={Fermi liquid instabilities in the spin channel},
  author={Wu, Congjun and Sun, Kai and Fradkin, Eduardo and Zhang, Shou-Cheng},
  journal={Physical Review B—Condensed Matter and Materials Physics},
  volume={75},
  number={11},
  pages={115103},
  year={2007},
  publisher={APS}
}

\clearpage
\section{Figures}
%\clearpage

\setcounter{figure}{0}
\makeatletter 
\renewcommand{\figurename}{Extended Data Figure}
\makeatother

\clearpage
\begin{figure}[H]
    \centering
    \includegraphics[width=1\textwidth]{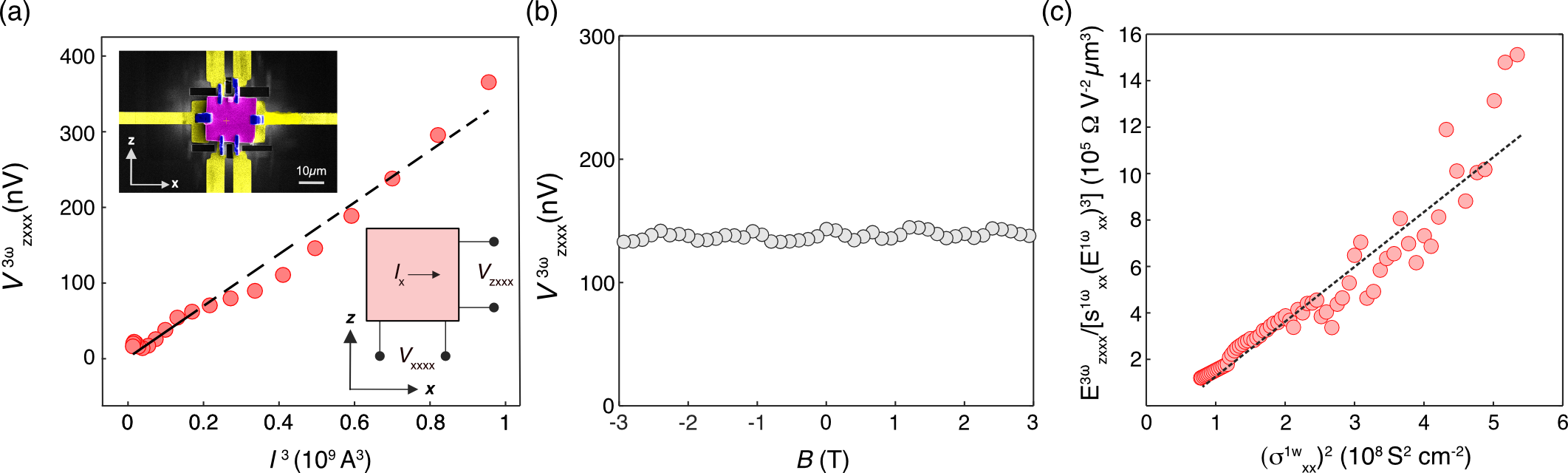}
    \caption{{\bf Reproduction of the third-order NLAHE on device $'XZ2'$.} We have fabricated another Hall bar device $'XZ2'$ from a CrSb lamella cut along the crystallographic $xz-$direction (see Methods section). (a) Shown is the third-order transverse voltage $V_{zxxx}^{3\omega}$ (circular symbols) recorded as a function of the cubed longitudinal bias current $I_x$ at room temperature $T=300\,$K. The measurement geometry and a false colored scanning electric microscopy image of the device are shown as an inset. The linear fit (dashed line) to the data establishes the third-order nature of the detected signal. The amplitude of $V_{zxxx}^{3\omega}$ measured in device $'XZ2'$ quantitatively reproduces the amplitude of $V_{zxxx}^{3\omega}$ measured in device $'XZ'$. (b) Shown is the magnetic field ($B$) dependence of $V_{zxxx}^{3\omega}$ at room temperature. Consistent with our observation on device $'XZ'$, $V_{zxxx}^{3\omega}$ is independent of the magnetic field amplitude over the examined field range. (c) Scaling law analysis of the third-order NLAHE. Shown is $E_{zxxx}^{3\omega}/(\sigma_{xx}^{1\omega}(E_{xx}^{1\omega})^3)$ (circular symbols) and linear fit to the data (dashed line) plotted as a function of the squared longitudinal conductivity $\left( \sigma_{\rm xx}^{1\omega} \right)^2$.From the fit, we extract a finite vertical intercept $\beta=(-106\pm3)\times10^3\,\Omega\mu\text{m}^3V^{-2}$ and linear slope $\alpha=(2381\pm16)\times10^{-6}\Omega^3\mu\text{m}^5V^{-2}$.}
    \label{fig:extfig1}
\end{figure}

\begin{figure}[H]
    \centering
    \includegraphics[width=0.5\linewidth]{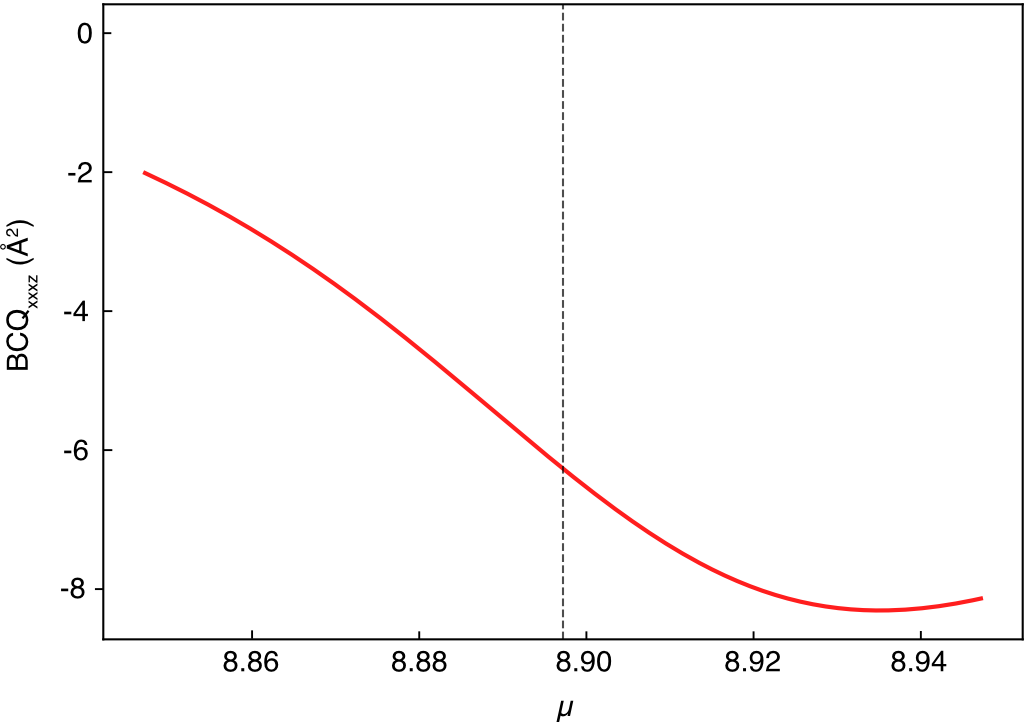}
    \caption{{\bf Finite Berry curvature quadrupole at Fermi energy from DFT calculations.} Shown is the momentum-space integrated Berry curvature quadrupole $BCQ_{xxxz}$ plotted as a function of the chemical potential $\mu$. The actual chemical potential of CrSb $\mu_0=8.8972 eV$ is indicated by a vertical dashed line. The underlying Berry curvature distribution $\mathcal{Q}_{xxxz}(k_x,\,k_z)$ in the $xz$-plane of the Brillouin zone is shown in Fig.~\ref{fig:fig3}(e). Details of the model calculation are presented in the Methods section.}
    \label{fig:extfig2}
\end{figure}

\clearpage
\section{Methods}

\subsection{Crystal synthesis and characterization}
Bulk single crystals of CrSb were grown by a flux method as described in detail elsewhere~\cite{urata2024high}. The obtained crystals are of hexagonal rod-like shape with typical dimensions of $0.2\ \rm{mm} \times 2\ \rm{mm}$ along the short and long axes, respectively.
Out-of-plane X-ray diffraction and transmission Laue measurements confirmed the single-crystalline nature of the CrSb samples and revealed that the lateral facets correspond to the $\{01\bar{1}0\}$ planes, with the long axis parallel to the $c-$axis.

\subsection{Microstructured device fabrication using FIB}

Crystalline lamellas of CrSb for Hall bar device fabrication were prepared using an FEI Helios G4 UX Dual Beam FIB system. Thin lamellas of dimensions $20~\mu\text{m} \times 10~\mu\text{m} \times 2~\mu\text{m}$ were FIB-cut and extracted from the same piece of bulk CrSb single crystal along different crystallographic directions with an alignment accuracy of $0.5^\circ$ to the respective crystallographic planes. The lamellae were then transferred to a copper grid using a platinum micromanipulator to perform thinning of the lamella and to remove an amorphous layer resulting from the FIB-cutting process using a low-voltage beam shower. Six-terminal Hall bars of dimensions $15~\mu\text{m} \times 8~\mu\text{m}$ were prepared on SiO$_{2}$/Si wafers of size $1~\text{cm} \times 1~\text{cm}$ followed by photolithography and Ti(5~nm)/Au(50~nm) deposition using an e-beam evaporator. Each lamella was welded to the gold electrodes using Pt deposition with the FIB to prepare the final devices.

\subsection{Electric transport measurements}

All electric transport measurements were performed using a Quantum Design PPMS 6000 system using a four-probe contact geometry. The Hall bar devices were wire-bonded to the chip holder using 25~$\mu$m thick aluminum wire. An a.c. bias current of $I = I_{0} \sin(\omega t)$ with a frequency of $f = \omega/2\pi = 19.375$~Hz was applied by using an a.c. current source (6221A, Keithley). Unless otherwise noted, an amplitude of $I_0=0.6\,$mA was used. Three lock-in amplifiers (SRS 830, Stanford Research Systems), which are phase matched to the bias current output, were used to simultaneously record nonlinear longitudinal voltages $V_{\mathrm{\alpha\alpha}}^{1\omega}$, $V_{\mathrm{\alpha\alpha\alpha}}^{2\omega}$, $V_{\mathrm{\alpha\alpha\alpha\alpha}}^{3\omega}$. Similarly, nonlinear transverse voltages $V_{\mathrm{\beta\alpha}}^{1\omega}$, $V_{\mathrm{\beta\alpha\alpha}}^{2\omega}$, and $V_{\mathrm{\beta\alpha\alpha\alpha}}^{3\omega}$ were recorded simultaneously in a separate experimental run. Control experiments that confirm the accuracy of our lock-in detection scheme are presented in Sec.~V of the Suppl.~Materials.

\subsection{Angle-resolved photoelectron spectroscopy measurements}
Conventional ARPES measurements were performed at the beamline UE112 PGM-2b-$1^2$ of BESSY (Berlin Electron Storage Ring Society for Synchrotron Radiation) synchrotron. The energy and angular resolutions were set to $\approx 20$~meV and 0.3$^\circ$, respectively, and the temperature was set to $\approx 20$~K. The bulk crystalline samples for all ARPES measurements were cleaved \textit{in situ} and measured in a vacuum better than $2 \times 10^{-10}$~Torr.

\subsection{Density Functional Theory calculation}
Density functional theory (DFT) calculations of the electronic structure were performed using the density functional theory framework as implemented in the Vienna {\em ab initio} simulation package~\cite{kresse1996efficiency, kresse1996efficient}. The projector-augmented wave potential was adopted with the plane-wave energy cutoff set to 520\,eV (convergence criteria $10^{-6}\,$eV). The exchange-correlation functional of the Perdew–Burke–Ernzerhof type was used~\cite{perdew1996generalized} with a 16$\times$16$\times$16 gamma-centered Monkhorst–Pack mesh. We constructed the Hamiltonian of a Wannier tight-binding model using the {\em WANNIER90} interface~\cite{mostofi2008wannier90}, including the Cr $d$-and Sb $p$-orbitals. We then calculated the Berry curvature and quantum metric from this Wannier model using a 201$\times$201$\times$151 k-mesh.

\clearpage
\section{Acknowledgments}

The authors appreciate valuable discussions with Max Hirschberger and Shiming Lei. We also acknowledge the support of Siu Hin Kee for the sample fabrication using the focused ion beam method at the Materials Characterization and Preparation Facility at HKUST. This work was primarily supported by the Hong Kong Research Grant Council (Grant Nos.\,26304221, 16302422, 16302624 awarded to BJ and C6033-22G awarded to BJ, JM, and JL), the Croucher Foundation (Grant No.\,CIA22SC02 awarded to BJ), and the National Key R$\&$D Program of China (Grant No.\,2021YFA1401500 awarded to JL). JL further acknowledges support from the Hong Kong Research Grants Council (Grant Nos.\,16306722, 16304523, and C6046-24G). JM was supported by the National Natural Science Foundation of China (Grant No. 12422405), the Hong Kong Research Grants Council (Grant Nos. 21304023, C6033-22G, C1018-22E, C6046-24G, and C1002-24Y).
TU was supported by Foundation of Public Interest of Tatematsu.

\section{Author Contributions}

BJ, JL, and SS conceived the project. SS fabricated the FIB-cut Hall bar devices samples and conducted the electric transport measurements. Xingkai C, Xinyu C, and FX conducted the symmetry analysis and the model calculations and analyzed the data together with SS. Single crystals were grown by TU, WH, HI, CD, and CF. BJ, JL, and JM supervised the study. BJ wrote the manuscript with the input of all authors.

\section{Competing Interest Declaration} The authors declare that they have no competing financial interest.

\section{Data Availability Statement} Replication data for this study can be accessed on Zenodo via the link XXX.

\end{document}